\documentclass[12pt]{article}
\def\be{\begin{equation}} \def\ee{\end{equation}}
\def\bi{\begin{itemize}} \def\ei{\end{itemize}}
\def\bea{\begin{eqnarray}} \def\eea{\end{eqnarray}} \def\ba{\begin{array}}
\def\ea{\end{array}} \def\ben{\begin{enumerate}} \def\een{\end{enumerate}}

\newcommand{\eqn}[1]{(\ref{#1})}

\newcommand{\hepth}[1]{{\tt [arXiv:{#1} [hep-th]]}}

\def\br{\nonumber\\}

\def\G{\Gamma}
\def\tr{{\rm Tr}}

\begin{document}
{}~
\hfill\vbox{\hbox{hep-th/1211.3281} \hbox{\today}}\break

\vskip 2.5cm
\centerline{\large \bf
The Yang-Mills and chiral fields in six dimensions 
}
\vskip .5cm

\vspace*{.5cm}

\centerline{  
 Harvendra Singh 
}
\vspace*{.25cm}
\centerline{ \it  Theory Division, Saha Institute of Nuclear Physics} 
\centerline{ \it  1/AF Bidhannagar, Kolkata 700064, India}
\vspace*{.25cm}
\vspace*{.25cm}

\vspace*{.5cm}

\vskip.5cm
\centerline{E-mail: h.singh (AT) saha.ac.in }

\vskip1.5cm

\centerline{\bf Abstract} \bigskip
In  previous work \cite{hs11},  we constructed an 
action in six dimensions using Yang-Mills fields and an auxiliary
Abelian field.  Here we first write down all the  equations
of motion and the constraints which arise
from such an action.  From these equations we reproduce
all dynamical equations and the  constraints required for 
 self-dual tensor field theory constructed by Lambert-Papageorgakis,
which describes  (2,0) supersymmetric CFT in 6D. This is 
an indication of the fact that our 6D gauge theory contains
all the same information as the on-shell theory of 
chiral tensor fields.

\vfill 
\eject

\baselineskip=16.2pt


\section{Introduction}
Recent progress of   holographic
membrane theories \cite{blg}-\cite{hs}, 
 provides us with ample motivation to try and understand  the 
 6-dimensional M5-brane theory. The latest attempts on this subject
can be found in the works \cite{lp,doug10,lps}.
While some headways in constructing such theories
having maximal supersymmetry could be found in the papers
\cite{lp,hs11}, 
and the subsequent generalisations in 
\cite{Samtleben:2011}-\cite{Dolan:2012wq}. 
As per the current understanding, the 
dynamics of single M5-brane is governed by an Abelian
 6D conformal tensor theory
having  maximal (2,0) supersymmetry. The antisymmetric
2-rank tensor fields are  natural  to occur 
in six dimensions. There are other important dynamical
 reasons to include tensors in these 6D constructions. 
Let us take the example of an extended M2-brane ending on 
 M5-brane. The intersection of these extended branes 
produces an infinitely long line defect on the world-volume of 
M5-brane. Such  defects do constitute the simplest excitations which
entirely live on the M5-brane. Basically, the defects 
 behave  like  extended `strings' living in a six-dimensional flat spacetime. 
It also makes us  believe  that ultimately the 
dynamics of these  stretched string-defects will
constitute the low energy  dynamics of the M5-branes. We may also consider 
 other configurations where  we have $N$  parallel (coincident) M5-branes
and a single M2-brane ends on them. In that situation M2-brane will 
produce line
defects on each single M5-brane in the stack. 
Thus we will have a lowest energy 
configuration on the stack  which has to be described by 
$N$ parallel (spatially aligned) 
strings in 6D. Of course, these `lowest' energy configurations 
would spontaneously break  the rotational symmetry on the 5-branes
from $SO(5) \to SO(4)$. 
\footnote{The situation here may crudely be compared to
 the case of alignment of spins in magnetism 
in the low energy (temperature) states. Full rotational symmetry in these
systems is obtained 
only in the disordered (high temerature) phase.} Thus we see that
 low energy states (vacua) of  M5-brane theory 
could well have manifestly broken Lorentzian symmetry. Hence
it would be worth while to include auxilary Abelian vector, $\eta^M$, 
in the  6D gauge 
theory to describe this low energy dynamics, so long as Lorentz invariant 
configurations (vacua) are also permitted in the theory. 
It is known  that
the v.e.v. of this auxiliary vector field 
will always break the Lorentz symmetry. 

The dynamical strings  would naturally couple to 
antisymmetric tensor field, $B_{MN}$, whose field strength  
$H_{(3)}=dB_{(2)}$ is a 3-form. But this field strength needs to be self-dual
in order to describe M5-brane. The   string like  solutions living
 on  M5-brane are already known to exist \cite{howe}. 
In fact, a self-dual  tensor field, 
{\it five} scalars, $X^I$, and  a Majorana-Weyl
 spinor, $\Psi$, constitute what is known as the simplest
 (2,0) tensor multiplet in 6-dimensions \cite{howe1}. 
The dynamical equations of  chiral tensor theory  are 
\bea
\label{self1a}
H_{(3)}\equiv dB_{(2)}=\star_6 H_{(3)}, ~~~
\partial_M\partial^M X^I=0=\not\!\partial\Psi
\eea
where $\star_6$ is the Hodge-dual  in six dimensions. 
This Abelian tensor theory   
is  superconformal,  but  the theory is trivial
as it is not interacting. It is  being currently argued that
all the states of a non-abelian (2,0) tensor theory, when compactified
on a circle, are perhaps contained in the 5-dimensional 
super-Yang-Mills (SYM) theory. As such  5D SYM is known to be
 nonrenormalizable and has a  strongly coupled fixed point in the UV. 
But if  SYM indeed contains all the states of a compactified 6D CFT
without requiring  new  degrees of freedom at higher loops, then 
the SYM ought to be be a finite theory in itself \cite{doug10,lps}. 
Although intuitive, but it is a very difficult 
to directly task to check  the  finiteness of 5D SYM. Any deviation from
the expected  behaviour of SYM will have direct 
consequences for 6D (2,0) theory, 
see recent attempts in this direction \cite{doug12}. 

Although
very little is known about the `non-Abelian' 
(2,0) tensor theory, which is supposed to  
describe the  dynamics on the stack of M5-branes, but
some attempts have been made recently 
to write down a theory using self-dual tensors \cite{lp}, 
and by directly uplifting 5D Super-Yang-Mills action to 
six-dimensions \cite{hs11}. Actually,
a non-Abelian 6D CFT, in a simple setting, should 
possess  $SU(N)$  gauge symmetry and  $SO(5)$  global symmetry as well as 
conformal symmetry. The 6D gauge action provided in \cite{hs11} inherits
 some of these features directly from  SYM, as it is a direct uplift
from 5D. Nevertheless these 
are some of the  requirements which may guide  us in the construction of
a meaningful  M5-brane theory.  
\footnote{ See  earlier developments on M5-brane  
in the references 
\cite{witt,ps,tonin,howe2,Howe:1997fb,Lee:2000kc,tonin1}.}

The goal of this work is to present a 6D  action involving Yang-Mills fields,
and an auxiliary vector  field following our
 earlier work \cite{hs11}. We
 write down all the equations of motion of this  theory determined 
by its action. 
We  then 
 show that these equations are the same as in the work of 
Lambert-papageorgakis \cite{lp}, which involves an on-shell construction of
(2,0) chiral tensor theory. The paper is organised as follows. 
In section-2, we systematically 
work out the equations of motions for the Abelian and non-Abelian theories
and also write down the constraint equations in these theory.
We then introduce self-dual tensor fields and rewrite field equations
in terms of these chiral tensors. In section-3, we present some solutions of 
the theory. The conclusions are given in the section-4.

\section{6D  gauge field theories}

\subsection{Abelian  gauge fields and chiral fields}
It has been proposed recently \cite{hs11} that
a covariant six-dimensional gauge action (in an axial form) 
involving scalar fields,
  could be written as
\bea\label{act3}
S&\equiv&\int d^6x  \bigg[
-{1\over 12 (\eta)^4} ( G_{MNP})^2  
 -{1\over2} (\partial_M X^I)^2 
  \bigg]
\eea
where $G_{MNP}$ itself is
of Chern-Simons type
 \be
G_{MNP}=\eta_{M}F_{NP}+ ~{\rm cyclic~ permutations~of~indices}
\ee
while  gauge field strength 
 $F_2=d A_1$. The vector $\eta^M$ will be taken to be
 constant everywhere, i.e. $d\eta=0$, but it  could be lifted 
to be a proper abelian  
field with the help of a Lagrange multiplier  \cite{hs11}. 
The  $X^I$'s $(I=6,7,...,10)$ are  five  real scalar fields.
Note that the gauge kinetic term in the action \eqn{act3} 
is rather of unusual type. But this  axial 
form  of  gauge  action helps us in working with reduced
 gauge degrees of freedom  (namely 3 on-shell vector
 d.o.f.s in 6D) in this special kind of covariant theory. 
The  equations of motion following from the above action can  be written as
\bea\label{hl1}
&& \partial_M \partial^M X^I=0,~~~~~~d\eta=0
 \\
&& \eta\wedge d\star G_3=0\ ,
\eea
Since $dF_2=d(dA)=0$, we can also write the Bianchi identity  as 
\be dG_3=0\ . 
\ee 
In our notation
 $\star$ is a Hodge-dual operation in a six-dimensional Minkowski 
space. The equations of motion
are all covariant and directly 
obtainable from the  action \eqn{act3}. 
Let us now consider some important contractions  involving
constant vector $\eta\equiv \eta_M dx^M$. It  
simply  follows  from the Bianchi,  $dF=0$, that the contraction
$\eta.(dF)=0$, which means that the following gauge identities involving
$\eta$ contractions   
\bea\label{ko8}
\eta^M\partial_M F_{PQ}= 0 =\eta^M F_{MN}
\eea
shall hold good. These equations are the nontrivial constraints and
would  remain implicit in our theory
with the Lagrangian given as in \eqn{act3}. Naturally, the theory 
will allow variety of solutions, 
$e.g.$ string-like  extended solutions, monoples and gauge instantons \cite{hs11}. 
One can find  other solutions too.
Thus, any given solution of the bosonic
equations will be characterised by namely 
the choice of $\eta_M,~ A_M$ and $X^I$. 
We would like to show that 
 the above equations, although  looking quite different,  
indeed  describe a  chiral field theory
 involving  self-dual 3-form tensors too!

\noindent{\bf Self-dual tensor fields:}
It can be noted that
we have not used any 2-rank 
anti-symmetric tensor field in the action \eqn{act3}. 
 However,
given the above set up, our next aim is to define a 3-form tensor, 
such that it is  consistent with the above equations of motion 
including the constraints described above and is also (anti)self-dual
in nature. Such 
a tensor field strength could be explicitly constructed out of $\eta$
and $F_2$ and it is given by
 \bea\label{hl5}
H_3 &\equiv& 
{1\over 2(\eta)^2}(\eta\wedge F +\star(\eta \wedge F)) \ .
\eea
It immediately follows from the dynamical equations \eqn{hl1} that 
$H$  satisfies the equation
\bea \label{hg3}
 dH={1\over 2(\eta)^2}d(\eta\wedge F +\star(\eta \wedge F))
=0\ . 
\eea
Thus given that  $\eta$ and $F$ being nontrivial, the tensor
 $H$ can always be introduced. Also
  by construction it will also be self-dual,
\be
 H=\star H\ .
\ee
In the next step, we  invert \eqn{hl5} and instead 
write down $F_2$ in terms of the contractions of
$\eta$ and $H$, whence
\bea
F_2= 2(\eta.H)\ .
\eea
From this  contraction we  get the identity
\bea\label{hg4}
dF=0=d(\eta.H)
\eea
Using eq.\eqn{hg3} we get the  constraint involving the tensor
\be\label{hg5}
\eta^M\partial_M H_{PQR}=0\ .
\ee

Actually we have taken up this exercise in order to relate our Yang-Mills 
field equations with those of
Lambert-Papgeorgakis (LP) involving self-dual tensors \cite{lp}. Indeed,
the bosonic  equations \eqn{hl1} and \eqn{hg3} \& \eqn{hg5} form the basis of 
(2,0) tensor field theory  
proposed by LP. Let us recall that
the LP  proposal had been solely based upon 
equations of motion, because there wouldn't
exist an action in 6D, directly involving self-dual tensors. 
However the  gauge action 
\eqn{act3} (albeit in the axial-form)
does the needful job  efficiently well. This leeway to have an action
is partly attached to the presence of auxiliary 
vector $\eta_M$ in our construction. Secondly, the action \eqn{act3} employs 
gauge fields  as fundamental dynamical entities 
and not the tensor fields. The tensor field  $H$ introduced in \eqn{hl5}
 in that case 
is merely a {\it composite} field.  

\noindent{\bf Including fermions:} So far we did not say anything about the 
fermionic fields. It would be interesting to 
include suitable fermionic fields in the action \eqn{act3}. Particularly,
the fermionic equation required for the on-shell
(2,0) supersymmetry \cite{lp} is 
\be\label{hg6}
\not\!\partial\Psi=0\ .
\ee
Thus a fermionic kinetic term  such as
$ \bar\Psi\not\!\partial\Psi$ needs to be   
added to the bosonic action \eqn{act3}.
The Abelian action including fermions becomes 
\bea\label{act3s}
S[A,X^I,\Psi]&\equiv&\int d^6x  \bigg[
-{1\over 12 (\eta)^4} ( G_{MNP})^2  
 -{1\over2} (\partial_M X^I)^2 
+{i\over 2}\bar\Psi\not\!\partial\Psi
  \bigg]
\eea
This action was originally proposed  in \cite{hs11}.
Importantly,  as we can see  here that
the eqs. \eqn{hl1},\eqn{hg3},  \eqn{hg6} as well as
the constraint \eqn{hg5}  do all follow from the action \eqn{act3s}. These 
equations  are those which  
describe  on-shell (2,0) supersymmetric theory \cite{lp}. 
The invariance of action \eqn{act3s} under supersymmetry 
\bea\label{susy1}
&& \delta_s X^I= i\bar\epsilon \Gamma^I\Psi\br
&& \delta_s A_{M}= i\eta^N\bar\epsilon \Gamma_{MN}\Psi\br
&& \delta_s \Psi= {1\over3!} H_{MNP}\Gamma^{MNP}\epsilon
+\partial_M X^I\Gamma^M \Gamma^{I}\epsilon \br
&&\delta_s\eta_M=0
\eea
will  however require other two
 constraints, namely
\be\label{vh2}
\eta^M \partial_M \Psi=0= \eta^M\partial_M X^I\ .
\ee
 (All spinors  have
32 real components. The  constant  spinors in supersymmetry transformations
satisfy the projection condition $\Gamma_{012345}\epsilon=\epsilon$.)
These latter constraints are the 
reflection of the fact that, although our Lagrangian density \eqn{act3s}
is superficially 6-dimensional, actual
on-shell dynamics of the fields lives in  5-dimensional space
only. We comment that
the constraints \eqn{vh2} cannot be derived from the Abelian
 action \eqn{act3s} due 
to the triviality (noninteracting nature)
of the theory,  
until unless we demand the closure of the action \eqn{act3s} under susy.
But these constraints will indeed follow rather simply
 in a non-Abelian (interacting) setting next. 

\subsection{  Non-Abelian chiral fields}
In the previous Abelian example we  learnt that it is possible to 
construct a gauge action in 6D, which reproduces the field equations of a
self-dual tensor theory. We would like to see if the same thing 
happens in the non-Abelian theory.
A  6-dimensional non-Abelian
gauge action including the fermions  could  be written as \cite{hs11} 
\bea\label{act3a}
S_{non-Abelian}&\equiv&\int d^6x \tr \bigg[
-{1\over 12 \eta^4} ( \eta_{[M}F_{NP]})^2  
 -{1\over2} (D_M X^I)^2 +{1\over4}(\eta)^2( [X^I,X^J])^2
 \br
&&~~+{i\over 2}\bar\Psi\Gamma^MD_M\Psi
-{1\over 2}\eta_M\bar\Psi\Gamma^M\Gamma^I[X^I,\Psi]
  \bigg]
\eea
where 
$F_{MN}=\partial_{[M} A_{N]} -i [A_{M}, A_{N}]$ 
is  the Yang-Mills field strength.
The  scalar fields $X^I$'s $~(I=6,7,8,9,10)$ are  also 
in the adjoint representation of the $SU(N)$.
The gauge covariant derivatives are 
\bea
&& D_M X^I= \partial_M X^I -i[A_M,X^I],~~~ 
D_M \Psi= \partial_M \Psi -i[A_M,\Psi] .
\eea
The $SU(N)$   gauge symmetry of the  
  action \eqn{act3a}  corresponds to the 
fact that there are $N$ parallel M5-branes.
The  gauge transformations are
\bea\label{gt1}
&&  A_{M}\to A'_{M}=U^{-1}A_{M} U - i U^{-1}\partial_{M} U\br
&&  X^I\to X'^I=U^{-1} X^I U, ~~~  
\Psi\to \Psi'=U^{-1} \Psi U
\eea
under which the action \eqn{act3a} 
remains invariant, where $U$ is an element of
$SU(N)$. 
We now study the equations of motion which follow from 
the  action \eqn{act3a}. Let us simplify our notation a bit and write 
the 2-form gauge field strength as
\bea
F_2\equiv D A= dA - i[A,A]
\eea
where  $D *=d * - i [A , *]$ is used for covariant derivative. 
The Bianchi identity for the Yang-Mills field is  
\be DF=0\ .
\ee
 Since $\eta^M$ is a covariantly constant (Abelian) vector,
we would have
\bea\label{gj1} 
\eta\wedge DF=0, ~~~ {\rm or} ~~~ D(\eta\wedge F)=0. 
\eea 
Also the contraction  $\eta.DF$ would then 
imply the following constraints
\bea 
\eta^M D_M F_{PQ}^a=0, ~~~~~ 
\eta^M F_{MQ}^a=0  
\eea
where $a$ runs over adjoint representation of the gauge group.

Let us switch off the fermions initially. 
The gauge field equations obtained from the action \eqn{act3a} are
\bea\label{gjj2}
\eta\wedge 
D\star(\eta\wedge F^a)-\star (\eta)^4 (X^I_b DX^I_c)f^{abc}=0
\eea
Combining \eqn{gj1} and \eqn{gjj2}, it also implies that
\bea\label{gj2}
\eta\wedge D(\star\eta\wedge F^a+\eta\wedge F^a)- \star
(\eta)^4(X^I_b DX^I_c)f^{abc}=0\ .
\eea
At this stage, let us
 introduce a non-Abelian 3-form tensor, namely
\bea\label{hl5s}
H_3^a\equiv {1\over 2(\eta)^2}(\eta\wedge F^a +\star(\eta \wedge F^a))
\eea
in the same way as in the Abelian case. It is also 
 self-dual by construction. By
inverting \eqn{hl5s} we can also write down $F$ in terms of $H$,
\bea
F^a=2(\eta.H^a)
\eea
where we  used  the constraint  $\eta.F=0$. The gauge Bianchi
 $DF=0$,  implies that
\be
D(\eta.H)=0\ .
\ee
It now follows from \eqn{gj2} that the tensor $H_3$ satisfies an equation
\be \label{gj3}
\eta\wedge DH^a - {1\over2}\star(\eta)^2 (X^{Ib} DX^{Ic})f^{abc}=0.
\ee
From here it is straight forward to check
 that by taking a contraction of equation \eqn{gj3} 
with  $\eta$, this equation can also be rewritten as a Bianchi
\be \label{jk45}
 DH^a +  {i\over2} \eta.( \star X^{Ib} DX^{Ic})f^{abc}=0
\ee
with the constraint 
\be
\eta^M D_M H_{PQR}^a=0\ .
\ee
As an independent check once structure constants $f^{abc}$ vanish, i.e. 
for $U(1)$ case, eq.\eqn{jk45} immidiately reduce to the Abelian theory
of the last section. But
 in the $SU(N)$ case, eq.\eqn{gj3} further implies a constraint, namely
\be\label{cons1}
\eta^M D_M X^{I}=0\ .
\ee
For convenience, in standard tensorial notation,
eq.\eqn{jk45} would give
\be \label{jk46}
 D_{[M}H^a_{PQR]} -  {1\over2} f^{abc} \epsilon_{_{MPQRNS}}
\eta^{N}X^{Ib} D^S X^{Ic}=0.
\ee
The last equation is the same equation as obtained by Lambert-Papageorgakis,
when the tri-algebra there
has been reduced to an ordinary Lie-algebra.
The $X^I$ equations of motion obtained from the action \eqn{act3a} are
\bea\label{jk47}
D\wedge\star DX^I+\star(\eta)^2 [X^J,[X^I,X^J]]=0\ .
\eea
Including the fermions, the field equations become
\be \label{jk45f}
 D_{[M}H_{PQR]} +  {i\over2}  
\epsilon_{_{MPQRNS}}
\eta^{N}[X^{I}, D^S X^{I}]
- {1\over4}\epsilon_{_{MPQRNS}}
 \eta^N [\bar\Psi, \Gamma^S\Psi]=0
\ee
along with the constraint 
\be\label{cons0}
\eta^M D_M H_{PQR}^a=0\ .
\ee
and
\be\label{cons2}
\eta_M f_{abc}\bar\Psi^b \Gamma^M \Psi^c=0\ .
\ee
The last fermionic constraint implies that the inner product of fermionic
current with vector $\eta_M$ always vanishes in the vacuum. 
\footnote{ In a given vacua, if 
$\eta^M=(0,0,0,0,0,\eta^5)$ is aligned 
to be  along the $x^5$ direction, then the  5-th component of  6D 
fermionic current, namely $<[\bar\Psi,\Gamma^5\Psi]>$, would  vanish! 
It may look weird, but it is consistent with the prospect that
 we would like to obtain 5D SYM theory after reduction of the 6D theory 
on $S^1$. The
 5D SYM theory does not allow any operator such as 
  $[\bar\Psi,\Gamma^5\Psi]$.}

Finally, the equations of motion of $X^I$ and $\Psi$  are 
\bea\label{jk47f}
&&
D\wedge\star D X^I+\star(\eta)^2 [X^J,[X^I,X^J]]
+{1\over 2}\star [\bar\Psi,\not\!\eta\Gamma^I\Psi]=0\ ,
\\ &&
\not\!D\Psi+i\not\!\eta[X^I,\Gamma^I\Psi]=0
 \eea
repectively. Thus, what has been discussed so far
 follows mainly from the equations and constraints 
directly obtainable from the action \eqn{act3a}.
The constraint which does not seem to immediately 
follow from the above set of
 equations is
\be\label{jk47g}
\eta^M D_M\Psi=0
\ee
However, it is not difficult to figure out that eq.\eqn{jk47f}
will be consistent only when eq.\eqn{jk47g} is included as a constraint.
To ascertain this let us act with  the 
operator $\eta^M D_M$ on the equation
\eqn{jk47f} from the left. Using the constraint \eqn{cons1} we  find that
all terms except the fermionic term 
$\eta^M D_M (\eta_N [\bar\Psi,\Gamma^N\Gamma^I\Psi])$ do indeed vanish. Hence 
for the  equation \eqn{jk47f} 
to be consistent, the constraint \eqn{jk47g} must  follow.
In summary, we have obtained all the equations
and constraints, involving self-dual tensor field,
which describe (2,0) supersymmetry and these all follow from the action
\eqn{act3a}. Note that we did not require any supersymmetry  
arguments in the above, but 
whatever we have obtained in the form of the equations already describes a 
maximally supersymmetric theory.
The  supersymmetry variations of the fields can  be written
 in the covariant form as \cite{hs11}
\bea\label{susy2}
&& \delta_s X^I= i\bar\epsilon \Gamma^I\Psi\br
&& \delta_s A_{M}= i\eta^N\bar\epsilon \Gamma_{MN}\Psi\br
&& \delta_s \Psi= {1\over3!} H_{MNP}\Gamma^{MNP}\epsilon
+D_M X^I\Gamma^M \Gamma^{I}\epsilon
-{i\over 2}\eta_M[X^I,X^J]\Gamma^{IJ}\G^M\epsilon \br
&&\delta_s\eta_M=0.
\eea
  These match with those in \cite{lp}, for an ordinary Lie-algebra,
 if we keep in mind our definition
of the self-dual tensor. There is no need to write a
separate susy transformation for
$H_{MNP}$ as it can be obtained from the variation of $A_M$. 

\subsection{5D SYM}

It is evident from  6D covariant
action \eqn{act3a} that the vector $\eta^M$ is only an auxiliary field
and the equations of motion always 
require it to take a constant value on-shell. Thus 
  $\eta^M$  inevitably
picks up a particular spatial direction in the vacuum and as
a result the off-shell $SO(1,5)$ symmetry gets spontaneously broken
down to $SO(1,4)$ Lorentz subgroup. 
Hence the on-shell dynamics of the 6D
fields will be exactly the same as that of 5D SYM fields. 
The details on the reduction of the 6D gauge action
to 5D SYM can be found in \cite{hs11}.  
This involves the vev $\eta^M=g \delta^{M}_5$, 
the radius of circle, $R_5$, on which 6D theory is compactified
 and a rescaling of the fields. For example,
the YM coupling constant has to  be defined as
\be
 (g_{_{YM}})^{2}\equiv {(g)^2\over R_5}.\label{jku2}\ee
Note that $g$ has the dimensions of length. On compactification only 
length scale available in the theory is the radius $R_5$. So 
we can naively take $g\simeq k R_5$, where $k$ is a 
dimensionless parameter. With this Eq.\eqn{jku2} can also be written as
\be
 (g_{_{YM}})^{2}\equiv {(k)^2 R_5}.\label{jku2a}\ee
This is an expected relation, as suggested by \cite{doug10,lps},
between the 5D Yang-Mills coupling constant and the 
radius of compactification of the sixth coordinate. 
The 5D scalars and the spinor $(\tilde X^I,~\tilde\Psi)$ (written with tilde 
 here so as to distinguish them from 6D fields) 
must be related to their 6D counterparts $(X^I,~\Psi)$ as
\be
\tilde X^I(x^\mu)=(R_5)^{1\over2}X^I(x^\mu)
, ~~~ \tilde \Psi(x^\mu)=(R_5)^{1\over2} \Psi(x^\mu),
\label{jku3}\ee
while gauge fields are related as
\be
\tilde A_\mu(x^\mu)=A_\mu(x^\mu), ~~~~
 A_5=0.
\label{jku4}\ee
Note that, the fields have no dynamics along $x^5$ (a natural
isometry direction), 
and the coordinates $x^\mu$'s span 5D Minkowski space. 
The action \eqn{act3a} would then reduce to the 
5D SYM action 
\bea\label{act3sym}
S_{YM}&=&\int d^5x \tr \bigg[
-{1\over 4 g_{_{YM}}^2}  (F_{\mu\nu})^2  
 -{1\over2} (D_\mu  X^I)^2 +{1\over4}(g_{_{YM}})^2( [X^I,X^J])^2
 \br
&&~~+{i\over 2}\bar\Psi\Gamma^\mu D_\mu\Psi
-{g_{_{YM}}\over 2}\bar\Psi\Gamma^5\Gamma^I[X^I,\Psi]
  \bigg]
\eea
where tilde over 5D fields has been dropped. The arbitrary (dimensionless)
parameter
$k$ in the expression \eqn{jku2a} is related to the following fact. There is
an special scaling of the 5D  theory
\bea
&& g_{_{YM}}\to {k}~  g_{_{YM}} \br
&& X^I\to {1\over {k}} X^I
, ~~~  \Psi \to {1\over {k}} \Psi,
\label{jku6}\eea
under which SYM action rescales as: $S_{YM}\to {1\over k^2} S_{YM}$. 
Thus taking different values of $k$, but
keeping the same compactification radius, would produce in general different 
SYM actions. But these actions would differ only upto an over all factor 
of ${1\over k^2}$. One can also set $k=1$ in \eqn{jku2}. 
   We  avoid further  repetitions here as details can be found in \cite{hs11}. 
To recall, in \cite{hs11} the 6D action \eqn{act3a} was constructed  as 
a direct uplift of the 5D SYM action, by taking the coupling constant 
to be an auxiliary vector field, as was the case with 
(2,0) tensor theory \cite{lp}.
 Thus in a sense  action \eqn{act3a}
can be viewed as a `comformal dressing' of 
the 5D SYM theory in one higher dimension.\footnote{The  terminology `conformal dressing'  
has been suggested by the anonymous referee and is quite appropriate here.} 
Generally, the guiding spirit behind our approach
 has been  similar to  the case of membranes or `D2 to D2' \cite{bobby}.
Particularly, the reduction of the 6D covariant equations
to 5D SYM, involving a tri-Lie-algebra set-up, is also 
outlined in \cite{lp}.\footnote{ Note that 
in order to connect to the work of \cite{lp}, 
 one must take the vev $g= R_5$ in \eqn{jku2}, so that
 $(g_{_{YM}})^2\sim R_5$.} 

\section{ Vacuas}
There exist a number of supersymmetric vacua in the 6D gauge theory, 
some of which have been described in
\cite{hs11}.
Let us note that all of these 6D solutions will have at least 
one isometry direction due to the nontrivial constant v.e.v. of $\eta_M$. 
It is  evident from the construction of the action
that there would be no stable point-like 
solutions in the 6D theory.
We now list some of the static vacua of the theory and find out the
components of tensor $H$.
\bi
\item  Let us first consider Lorentz symmetric vacua. 
It corresponds to taking
$\eta_M= {\rm constant}$ and $X^I=u^I$, with $u^I$'s being
   $N\times N$ diagonal constant
matrices \cite{hs11}. 
The Yang-Mills fields are vanishing for these solutions. 
These vacua are the maximally supersymmetric configurations and describe the 
moduli space corresponding to $N$  M5-branes placed on a 
flat 5-dimensional transverse space. 
However, there  exists  an unique $(\eta)^2 \to 0$  limit    
of these solutions, such that
 when this limit is taken, the  vacua will also preserve 
full $SO(1,5)$ Lorentz symmetry of the theory.\footnote{ 
As it is clear from the actions \eqn{act3s} and \eqn{act3a} that these
actions could also be written in terms of inverse vector
$\xi^M={\eta^M\over (\eta)^2}$. 
In that case we should be taking the limit, $(\xi)^2\to \infty$.} 
 These are the only vacuas which 
admit full Lorentzian symmetry. 
 
\item
We next consider  solitonic
configurations describing an  
extended M2-brane ending  on M5-brane \cite{hs11}. 
Consider the vacuum where  $\eta^M=(0,0,0,0,0,g)$, 
   aligned along $x^5$,
 which we take to be an isometry direction.
That is the soliton (string) is aligned along $x^5$. This configuration is
\bea\label{hl2} 
&& X^I(x^m)=\delta^{I10} \phi(x^m),~~~~(I=6,7,8,9,10) \br
&& F_{0m}=\pm g \partial_m\phi .
\eea
This configuration is a solution of equations \eqn{hl1} provided
\be 
\phi(x^m)=\phi_0+\sum_{i=1}^{p}{2 q_i\over|x-\zeta^i|^2}
\ee
where fields depend upon  world-volume coordinates $x^m$ ($m=1,2,3,4$)
except  $x^5$. Here $\phi_0$ is an arbitrary constant, while
  $\vec{\zeta}^i, q_i$ are the  parameters such as 
positions and charges of the $p$ solitons. 
The supersymmetry  is preserved when 
\be
(1\mp \Gamma^0\Gamma^5\Gamma^{10})\epsilon =0
\ee
Since  only one of the scalar fields, namely $X^{10}$, 
representing a transverse coordinate, $x^{10}$,
 has been excited, we  have a 
description in which  M2-brane,  extending along $x^5$-$x^{10}$ plane, ends 
on the M5-brane. The intersection is  along the common direction $x^5$. 
Such a solitonic excitation (the intersection)
will create a one-dimensional string defect on
 M5 world-volume. 
The electric field surrounding the string,  $E_m\equiv F_{0m}$,  will be
  peaked near its location at $\zeta^i$. 
For this solution we can now calculate the nonvanishing components 
of the 3-rank tensor, using \eqn{hl5},
\be
H_{50m}= {1\over 2} \partial_m\phi, ~~~
H_{mnp}={1\over 2} \epsilon_{mnpl50}G^{50l}= {1\over 2} 
\epsilon_{mnpl}\partial_l\phi
\ee
where $\epsilon_{mnpl}$ is Levi-Civita tensor in four dimensions. 
It shows that $H$ is self-dual.

\item  We next consider a magnetic monopole configuration 
 \cite{howe}. We  take $\eta^M$ aligned 
along $x^5$, as above, but
we  consider $x^4$ to be  another isometry direction. We denote
 the remaining three spatial coordinates  by
$x^a$, with index $a=1,2,3$. 
Over this 3-dimensional Euclidean sub-space 
we have a magnetic monopole solution given by
\be\label{mm1}
F_{ab}=\mp g \epsilon_{abc} \partial_c \phi, ~~~X^{10}(x^a)=\phi(x^a)=
\phi_0+\sum_i{2 p_i\over|x-\zeta^i|}  
\ee
which solves all the equations of motion in \eqn{hl1}. 
For the supersymmetry variations to vanish
we require following condition on the constant spinors
\be
(1\pm \Gamma^4\Gamma^0\Gamma^{10})\epsilon =0.
\ee
Thus the 6D Abelian gauge theory admits ${1\over 2}$-BPS  
monopole like solutions \cite{hs11}.  

Correspondingly an electric type solution living over 
this 3-dimensional Euclidean sub-space is simply
\be\label{el1}
F_{0a}=\mp g \partial_a \phi, ~~~X^{10}(x^a)=\phi(x^a)=
\phi_0+\sum_i{2 q_i\over|x-\zeta^i|}  
\ee
where we instead took $\eta_M=(0,0,0,0,g,0)$, i.e. here 
4th component of $\eta$ is nonvanishing.
In this case, for the supersymmetry 
we still require 
\be
(1\pm \Gamma^4\Gamma^0\Gamma^{10})\epsilon =0.
\ee
This suggests that, if the (2,0) theory is compactified on $T^2$, 
these electric and magnetic solutions of \eqn{el1} \& \eqn{mm1} 
would map into each other under 
the S-duality of $4D$ SYM theory, 
provided that 
$$\eta_4 \leftrightarrow\eta_5.$$ 
It means that two sides of  $T^2$ over which (2,0) gauge theory is 
compactified gets exchanged when we implement
 $4D$ S-duality. This establishes the conclusions in
 \cite{doug10}.  

A mixed electro-magnetic solutions can also be found
if  we let $\eta_M$ to be a generic vector living on
on $T^2$, spanning  $(x^4,x^5)$. The 
gauge field strength, $F$, should be taken to have mixed
components, 
$(F_e,F_m)$, over rest of the  coordinates patch $(x^0,x^1,x^2,x^3)$. 
The amount of supersymmetry will depend upon the choice of various 
parameters like the charges. 

\item
Interesting instantonic solutions
are found when we take $\eta_M$ to be a  vector
having components only along,  $x^0 $ and $x^5$. We shall again take  
$\eta_M=(0,0,0,0,0,g)$ for simplicity, as  a
 boost can generate other component $\eta_0$.
 The gauge field strength $F$ is taken to be Yang-Mills  self-dual 2-forms
living over the  Euclidean patch $(x^1,x^2,x^3,x^4)$. Accordingly
the $H$-tensor will be
\be 
 H_3={1\over 2g} dx^+\wedge(F_2 +\star_4 F_2)
\ee
where $x^\pm=(x^0\pm x^5)/2$. We see that $H$ 
is definitely self-dual and satisfies $dH=0=d\star H$. All $X^I$'s 
are taken constant diagonal matrices \cite{hs11}.

\ei

\section{Conclusion}
We have explicitly shown that the equations and the constraints
 which follow from  6-dimensional gauge field action 
are the same  as the `on-shell construction'
of  (2,0) supersymmetric chiral tensor theory 
by Lambert-Papageorgakis. The important point to note is that
 all these equations  follow from  
covariant 6D gauge action, in which 
 the  algebra is taken to be an ordinary Lie-algebra, for simplicity. 
We have demonstrated that  (anti)self-dual tensors can always 
be introduced in our equations of motion with out  
requirement of any additional fields
or any new algebraic structure, such as tri-algebra. 
However, there would always exist generic extensions of such theories 
to include tri-Lie-algebra \cite{lp}. 
In an interesting development, the authors in \cite{Samtleben:2011} 
 recently  presented a (1,0) supersymmetric Lagrangian theory 
in six dimensions. 
Thus it would be  worth while to check if our 6D gauge action 
could be embedded into some reduction of the (1,0) supersymmetric theory.

\vskip.5cm
\noindent{\it Acknowledgements:}\\
I would like to thank  
Anirban Basu,  Neil Lambert, Shiraz Minwalla, Sunil Mukhi
and especially  Henning Samtleben for helpful discussions
and for reading the draft.
I am  also grateful to  the organisers of the
Isaac Newton Institute, Cambridge  workshop programme- 
`Mathematics and Applications of Branes in String and M-theory' 
for the  kind hospitality 
  where part of this work was carried out.


\end{document}